\documentclass[twocolumn,showpacs,preprintnumbers,amsmath,amssymb,prl]{revtex4}

\usepackage{graphicx}
\usepackage{dcolumn}
\usepackage{bm}
\usepackage{color}
\begin{document}

\preprint{}

\title{Excitation and recombination dynamics of vacancy-related spin centers in silicon carbide}

\author{T.~C.~Hain$^{1}$}
\author{F.~Fuchs$^{2}$}
\author{V.~A.~Soltamov$^{3}$}
\author{P.~G.~Baranov$^{3,4}$}
\author{G.~V.~Astakhov$^{2}$} 
\email[E-mail:~]{astakhov@physik.uni-wuerzburg.de}
\author{T.~Hertel$^{1}$}
\author{V.~Dyakonov$^{2,5}$}
\email[E-mail:~]{dyakonov@physik.uni-wuerzburg.de}

\affiliation{$^1$Institute of Physical and Theoretical Chemistry, Julius-Maximilian University of W\"{u}rzburg, 97074 W\"{u}rzburg, Germany \\
$^2$Experimental Physics VI, Julius-Maximilian University of W\"{u}rzburg, 97074 W\"{u}rzburg, Germany \\
$^3$Ioffe Physical-Technical Institute, 194021 St.~Petersburg, Russia \\
$^4$St.~Petersburg State Polytechnical University, 195251 St.~Petersburg, Russia \\
$^5$Bavarian Center for Applied Energy Research (ZAE Bayern), 97074 W\"{u}rzburg, Germany}

\begin{abstract}
We generate silicon vacancy related defects in high-quality epitaxial silicon carbide layers by means of electron irradiation. By controlling the irradiation fluence, the defect concentration is varied over several orders of magnitude.  We establish the excitation profile for optical pumping of  these defects and evaluate the optimum excitation wavelength of $770 \, \mathrm{nm}$. We also measure the photoluminescence dynamics at room temperature and find a monoexponential decay with a characteristic lifetime of $6.1 \, \mathrm{ns}$. The integrated photoluminescence intensity depends linear on the excitation power density up to $20 \, \mathrm{kW / cm^2}$, indicating a relatively small absorption cross section of these defects. 
\end{abstract}

\date{\today}


\maketitle

Silicon carbide (SiC) has recently been suggested for deployment in various quantum technologies   \cite{Baranov:2005em,Weber:2010cn,Baranov:2011ib,Koehl:2011fv,Gali:2012fd,Riedel:2012jq,Baranov:2013gi,Castelletto:2013jj}. In particular, silicon vacancy ($\mathrm{V_{Si}}$) related defects can be used as coherently controllable \cite{Koehl:2011fv,Soltamov:2012ey} and selectively addressable \cite{Riedel:2012jq,Falk:2013jq} qubits, characterized by long spin coherence and spin-lattice relaxation times even at room temperature. These atomic-scale qubits are also attractive for local sensing of magnetic \cite{Kraus:2013vf} and  electric \cite{Falk:2013wx} fields as well as temperature \cite{Kraus:2013vf}. Another potential use is as a low-noise quantum microwave amplifier based on the maser effect \cite{Kraus:2013di}. Furthermore, SiC is a technologically friendly material, used in various devices (LED, MOSFETS, MEMS, sensors). This facilitates attractive opportunities for electrical control of these vacancy-related spin centers \cite{Fuchs:2013dz,Klimov:2013ua}.  

In all these appealing applications, an optical pumping scheme of the high-spin ground state of $\mathrm{V_{Si}}$-related defects  ($S = 1$ or $S = 3/2$)  \cite{Vainer:1981vj,Wimbauer:1997fj,Sorman:2000ij,Mizuochi:2002kl,Orlinski:2003dw,Ilyin:2006wb,Son:2006im,Kraus:2013di} is exploited. This pumping scheme is assumed to be similar to that of the nitrogen-vacancy (NV) defect in diamond \cite{Jelezko:2006jq}. In spite of this similarity, the optical properties of the $\mathrm{V_{Si}}$-related defects in SiC have not been measured in detail. It is worth noting that another vacancy centers in SiC, associated with the so-called AB lines, have been demonstrated to be very efficient single photon sources \cite{Castelletto:2013el}. In these single-defect experiments, the positively charged configuration of these centers has been investigated, which has low-spin ground state ($S = 1/2$). Therefore, it is unclear whether their quantum control is possible in the same manner as for the $\mathrm{V_{Si}}$-related defects in SiC or the NV defect in diamond. 

\begin{figure}[btp]
\includegraphics[width=.45\textwidth]{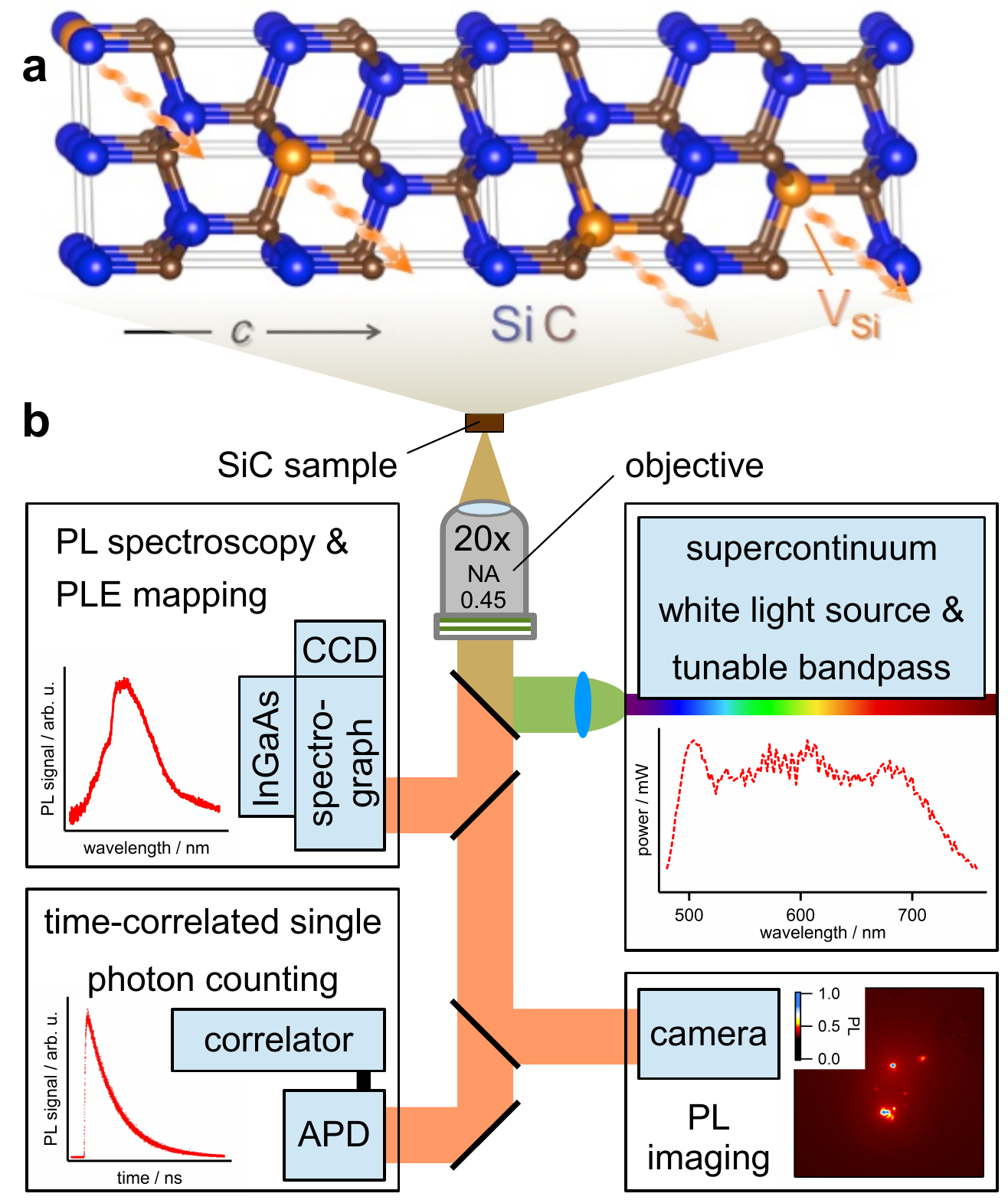}
\caption{(a) Crystal structure of 4H-SiC and silicon vacancy defects ($\mathrm{V_{Si}}$). The wavy arrows indicate the near infrared emission of $\mathrm{V_{Si}}$. (b) A scheme of the experimental setup for spatial imaging, spectroscopical and time-correlated measurements. } \label{fig1}
\end{figure}

In this Letter, we measure excitation spectrum and recombination dynamics of $\mathrm{V_{Si}}$ in 4H-SiC at room temperature.  The crystal structure of polytype 4H is shown in Fig.~\ref{fig1}(a). There are four Si and four C atoms per unit cell with stacking sequence ABCB (where A, B and C denote close-packed atomic planes). The broken bonds with missing Si atom result in the appearance of additional energy states and the electron density is mostly localized at the first- (C) and  second-neighbor (Si) atoms to this $\mathrm{V_{Si}}$ vacancy. According to the density functional theory (DFT) calculations \cite{Gali:2012jy}, the highest occupied state and the lowest (partially) unoccupied states lie deep in the forbidden gap of SiC, which is $E_g =  3.23 \, \mathrm{eV}$ at room temperature. For this reason, these defects can be detected by their characteristic photoluminescence (PL) in the spectral range shifted towards longer wavelengths with respect to the fundamental emission/absorption edge (ca. $400 \, \mathrm{nm}$ in 4H-SiC). At cryogenic temperatures, the so-called zero-phonon lines (ZPLs) are observed, distinct in their optical transition energies. In highly homogeneous samples, these ZPLs are very sharp with full width at half maximum (FWHM) of a few $\mathrm{\mu eV}$ (ref.~\onlinecite{Riedel:2012jq}). With rising temperature, the ZPLs are significantly broadened due to interaction with thermally excited phonons and their FWHM is approximately $100 \, \mathrm{meV}$ at $T = 300 \, \mathrm{K}$.

In 4H-SiC there are two non-equivalent lattice sites and, therefore, there are two different $\mathrm{V_{Si}}$ defects with corresponding ZPLs labeled as V1 and V2 (ref.~\onlinecite{Wagner:2000fj}). These defects are probably perturbed by carbon vacancies \cite{Baranov:2013gi}, which are not shown in Fig.~\ref{fig1}(a) for simplicity. When a silicon and carbon vacancies are situated along a chemical bond, a divacancy ($\mathrm{V_{Si}}$-$\mathrm{V_{C}}$) is formed \cite{Baranov:2005em,Son:2006im}, having its own ZPL. 

The 4H-SiC samples investigated in this Letter have been purchased from CREE. A nominally undoped (residual n-type doping $5.0 \times 10^{14} \, \mathrm{cm ^{-3}}$) layer of $110 \, \mathrm{\mu m}$ thickness was epitaxially grown on a 2-inch n-type 4H-SiC wafer ($396 \, \mathrm{\mu m}$ thickness, $7.84^{\circ}$ off-axis orientation). 
The layer is conserved by a 5-$ \mathrm{\mu m}$-thick n-type 4H-SiC layer ($1.2 \times 10^{16} \, \mathrm{cm ^{-3}} $) and a 2-$ \mathrm{\mu m}$-thick p-type 4H-SiC layer ($6.0 \times 10^{16} \, \mathrm{cm ^{-3}} $). The wafer was diced in  $4 \, \mathrm{mm} \times 2 \, \mathrm{mm}$ pieces. In order to create $\mathrm{V_{Si}}$-related defects the samples were irradiated with $0.9 \, \mathrm{MeV}$ electrons to different fluences. 

First, we present spatial imaging of defects in our SiC samples [our microscopy setup is schematically sketched in Fig.~\ref{fig1}(b)]. The sample is placed in an inverted Ti-U microscope and the PL is excited and collected through a 20x CFI Super Plan Fluor ELWD objective with $\mathrm{N.A.} = 0.45$ (Nikon). The excitation light from a SuperK EXTREME (EXR-15) supercontinuum laser (NKT Photonics) was defocused onto a spot of approximately  $50 \, \mathrm{\mu m}$ in diameter. The PL image in the confocal plane of the objective is recorded by a Clara interline CCD camera (Andor), with one CCD pixel ($6.45 \, \mathrm{\mu m}$) corresponding to $323 \, \mathrm{nm}$ at the sample position. The difraction-limited spatial resolution of the setup is about $1.3 \, \mathrm{\mu m}$. We use a wide-band excitation ($480 - 740 \, \mathrm{nm}$) for imaging, and the emission is detected for longer wavelengths ($\lambda > 760  \, \mathrm{nm}$).  

\begin{figure}[btp]
\includegraphics[width=.43\textwidth]{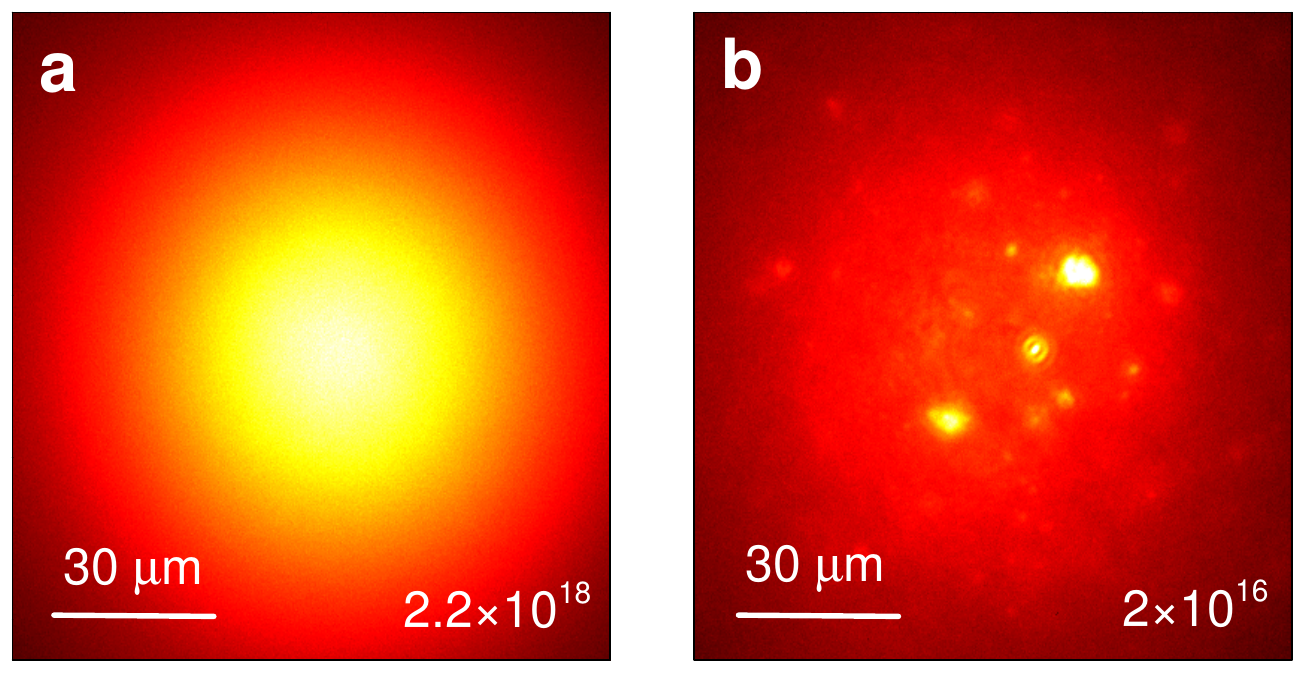}
\caption{(a) Spatial imaging of the $\mathrm{V_{Si}}$-related PL in a strongly irradiated sample (irradiation fluence $2.2 \times 10^{18} \, \mathrm{e / cm^{2}}$).  (b) The same as (a) but in a weakly irradiated sample (irradiation fluence $2 \times 10^{16} \, \mathrm{e / cm^{2}}$). Note, the intensity scales in (a) and (b) are different.} \label{fig2}
\end{figure}

To measure PL spectra we use a  Shamrock SR-303i spectrograph (Andor) together with either a Newton CCD camera (DU920P-OE) for detection from $760$ to $1000 \, \mathrm{nm}$ or a iDus InGaAs array detector (DU491A-1.7) for detection from $1000$ to $1500 \, \mathrm{nm}$ [Fig.~\ref{fig1}(b)]. This configuration provides the spectral resolution of ca. $0.6 \, \mathrm{nm}$.  For photoluminescence excitation (PLE) measurements, the laser wavelength is selected from the supercontinuum spectrum with a step of $2 \, \mathrm{nm}$ using a SuperK VARIA bandpass filter (NKT Photonics). In low-temperature experiments ($T = 10 \, \mathrm{K}$), the sample is mounted in a cold-finger MicrostatHe (Oxford Instruments) and placed in a LabRAM system for spectroscopy (Horiba Scientific). In this case we use a He-Ne laser ($633 \, \mathrm{nm}$) for excitation. 

Figure~\ref{fig2} shows PL images of two samples irradiated with electrons to different fluences. The wavelength range detected by the PL microscope setup is associated with emission from color centers (below we demonstrate that these colour centres are the $\mathrm{V_{Si}}$ defects). At a high irradiation fluence ($2.2 \times 10^{18} \, \mathrm{e / cm ^{2}}$), the PL intensity appears to closely follow the excitation profile which suggests that emission is from a homogeneous distribution of color centers [Fig.~\ref{fig2}(a)]. At a much lower irradiation fluence ($2 \times 10^{16} \, \mathrm{e / cm ^{2}}$) this profile splits up into several more strongly emissive fragments  [Fig.~\ref{fig2}(b)]. Some of these fragments have sub-$\mathrm{\mu m}$ size and reveal diffraction rings. Neither of these is observed in as-grown samples suggesting that these color centers were produced by electron irradiation. 

\begin{figure}[btp]
\includegraphics[width=.47\textwidth]{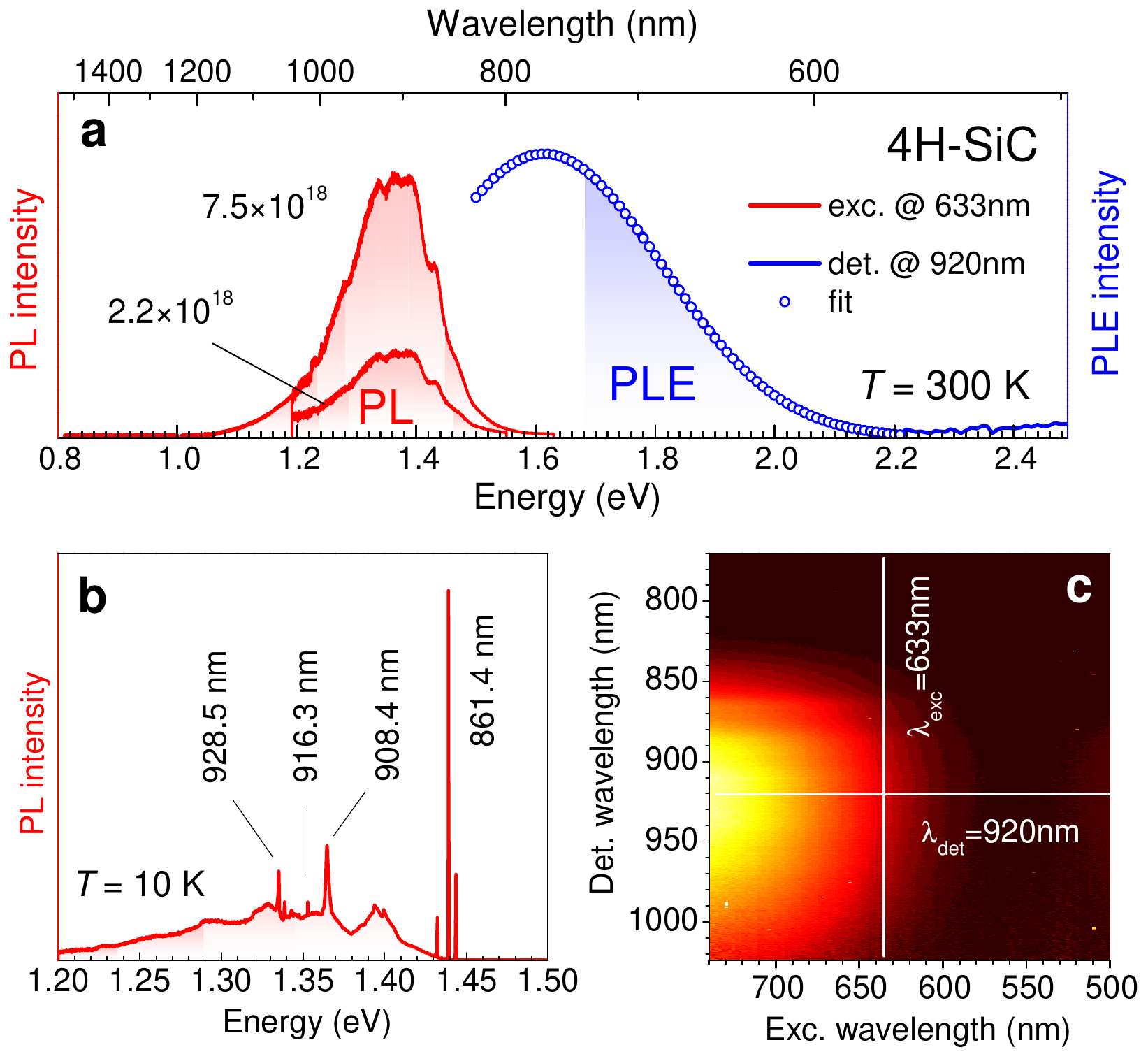}
\caption{(a) Room-temperature PL and PLE spectra of $\mathrm{V_{Si}}$ in 4H-SiC for different irradiation fluences. The Gaussian fit of the PLE spectrum is shown by open circles. (b) A low-temperature PL spectrum (irradiation fluence $7.5 \times 10^{18} \, \mathrm{e / cm^{2}}$).  The V1 and V2 ZPLs of $\mathrm{V_{Si}}$ are observed at $861.4 \, \mathrm{nm}$ and $916.3 \, \mathrm{nm}$, respectively. (c) A 2D scan of the excitation and detection wavelengths at room temperature. } \label{fig3}
\end{figure}

The room-temperature PL spectra are presented in Fig.~\ref{fig3}(a).  With increasing electron irradiation from $2.2 \times 10^{18} \, \mathrm{e / cm ^{2}}$ to $7.5 \times 10^{18} \, \mathrm{e / cm ^{2}}$ the integrated PL intensity increases ca. by a factor of three, suggesting that the defect concentration is proportional to the irradiation fluence. We observe PL in the spectral range from $850$ to $1050 \, \mathrm{nm}$, which is characteristic for $\mathrm{V_{Si}}$ in SiC at room temperature \cite{Fuchs:2013dz}. On the other hand, we do not observe PL in the spectral range from $1050$ to $1350 \, \mathrm{nm}$, where PL from $\mathrm{V_{Si}}$-$\mathrm{V_{C}}$ is expected \cite{Koehl:2011fv}. By cooling down to $T = 10 \, \mathrm{K}$ we resolve four ZPLs in the PL spectrum [Fig.~\ref{fig3}(b)]. The V1 ZPL at $\lambda = 861.4 \, \mathrm{nm}$ dominates in the spectrum, unambiguously proving that we generate and probe the $\mathrm{V_{Si}}$ defects. 

In oder to find the optimal excitation conditions we perform PLE mapping experiments. First, we ascertain that the PL spectrum is independent of the excitation wavelength [Fig.~\ref{fig3}(c)]. A PLE spectrum, detected at the PL maximum ($\lambda = 920 \, \mathrm{nm}$) and corrected for the laser supercontinuum spectrum, is presented in Fig.~\ref{fig3}(a). With decreasing wavelength ($\lambda < 760 \, \mathrm{nm}$) the excitation becomes less efficient and for $\lambda < 590 \, \mathrm{nm}$ the $\mathrm{V_{Si}}$ PL is not excited. The corresponding energy $2.1 \, \mathrm{eV}$ is still significantly smaller than the bandgap in 4H-SiC ($3.23 \, \mathrm{eV}$). According to the Franck-Condon principle, the PLE maximum is shifted towards higher energy with respect to the ZPL. In oder to estimate this maximum, we fit the PLE spectrum in Fig.~\ref{fig3}(a). In fact, the PLE spectrum associated with vibronic bands in the crystal has a complex form and is usually asymmetric with respect to the maximum. We find that the high-energy tail of the experimentally obtained PLE spectrum ($E > 1.68 \, \mathrm{eV}$) is perfectly described by the Gaussian in the form $\exp (-2 (\frac{E - E_{max}}{ \Delta})^2 )$ with $E_{max} = 1.61 \, \mathrm{eV}$  and $\Delta = 0.40 \, \mathrm{eV}$ [open circles in Fig.~\ref{fig3}(a)], indicating that inhomogeneous broadening dominates the excitation spectrum.  Following this fit, the optimal excitation wavelength is around $770 \, \mathrm{nm}$. Interestingly, we observe non-equivalent linewidth of the PL and PLE spectra, expected to be similar from a configuration coordinate diagram approach. A possible explanation could be due to the different interaction with phonons in the ground and excited states. Further understanding requires detailed theoretical consideration, which is beyond the scope of this work.

\begin{figure}[btp]
\includegraphics[width=.47\textwidth]{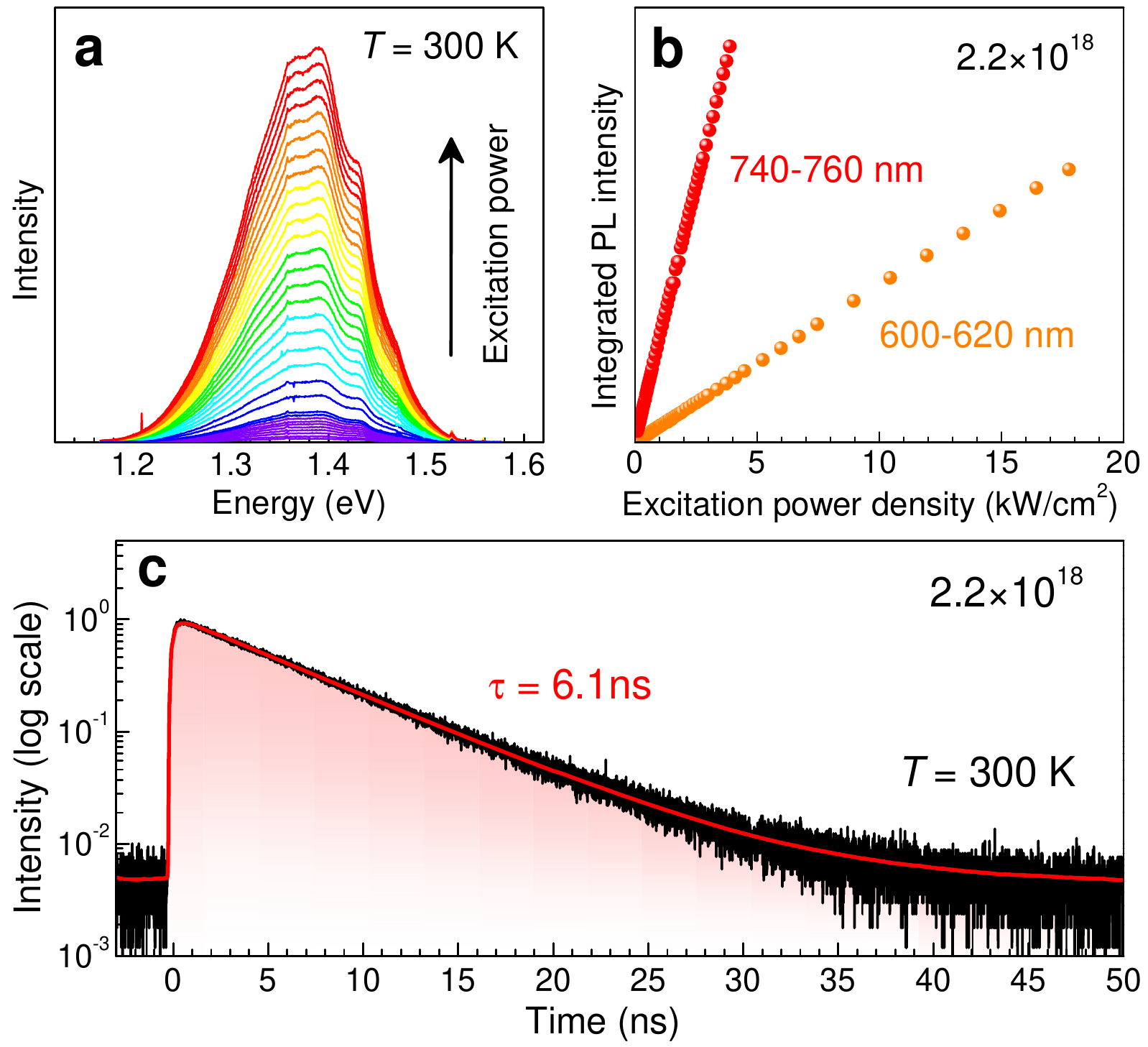}
\caption{(a) PL spectra with rising excitation power. (b) Excitation power density dependence of the integrated PL intensity for two different excitation spectral ranges. (c) Monoexponential PL decay of $\mathrm{V_{Si}}$ with a characteristic lifetime $\tau = 6.1 \, \mathrm{ns}$. } \label{fig4}
\end{figure}

To estimate the excitation efficiency we measure the integrated PL intensity as a function of excitation power [Fig.~\ref{fig4}(a)]. It is presented in Fig.~\ref{fig4}(b) for two spectrally different excitations. The excitation in the spectral range $740 - 760 \, \mathrm{nm}$ is more efficient than in $600 - 620 \, \mathrm{nm}$, as  expected from the PLE spectrum of Fig.~\ref{fig3}(a). We observe perfectly linear dependences on the excitation power density up to $20 \, \mathrm{kW / cm^2}$ ($600 - 620 \, \mathrm{nm}$) and $3 \, \mathrm{kW / cm^2}$ ($740 - 760 \, \mathrm{nm}$), indicating that all the results reported here are obtained far below the saturation. Remarkably, the saturation power density for the NV defect is about $100 \, \mathrm{kW / cm^2}$ (refs.~\onlinecite{Jelezko:2006jq, Kurtsiefer:2000tk}). 
As we do not observe deviation from the linear dependence, this suggests that the absorption cross section of the $\mathrm{V_{Si}}$ defects is comparable or smaller than that of the NV defect.  

Finally, we analyze PL dynamics of $\mathrm{V_{Si}}$, using time-correlated single photon counting (TCSPC) measurements. The repetition rate of the pulsed ($5 \, \mathrm{ps}$) excitation is $10 \, \mathrm{MHz}$ and the spectrally integrated PL signal is detected using a Si single-photon avalanche photodiode and a TCSPC HydraHarp module (PicoQuant) [Fig.~\ref{fig1}(b)]. To characterize our setup we measure the instrument response function, using scattered light from the laser in the spectral range from $800$ to $840 \, \mathrm{nm}$, and obtain the overall time resolution of $50 \, \mathrm{ps}$. 

The decay of the $\mathrm{V_{Si}}$ PL after pulsed excitation is presented in Fig.~\ref{fig4}(c). The experimental data are perfectly fitted with a monoexponential decay of the form $a + b \exp (-t / \tau)$ convoluted with the instrument response function. From the fit we find the PL lifetime $\tau = 6.1 \, \mathrm{ns}$. This is of an order of magnitude comparable to that of another color center in SiC ($1.2 \, \mathrm{ns}$)\cite{Castelletto:2013el} and of the NV defect in diamond ($11.7 \, \mathrm{ns}$)\cite{Kurtsiefer:2000tk}. This similarity is expected because all these atomic-scale defects are isolated from the bulk bands and their optical transitions are determined by the coupling of the electronic and vibronic states. The transition rate in this case, as a rule, is smaller than for direct band-to-band transitions in semiconductor nanostructures, like colloidal quantum dots \cite{Klimov:2000go} and quantum wells \cite{Yakovlev:2000ez}.  

Summarizing, we have performed spatial imaging of the $\mathrm{V_{Si}}$ defects, generated by electron irradiation in epitaxial 4H-SiC layer, and investigated their optical properties. In particular, we have found the optimal excitation conditions to optically pump these defects and we have measured their PL lifetime. This information is important to optimize quantum control of these defects, as required for various appealing applications. 

This work has been supported by the Bavarian Ministry of Economic Affairs, Infrastructure, Transport and Technology as well as  by the DFG under grant AS310/4. 



\end{document}